\documentclass[aps,pr,superscriptaddress,preprintnumbers,nofootinbib,10pt]{revtex4-2}

\usepackage{multirow}
\usepackage{amsmath}
\usepackage{amssymb}
\usepackage[dvipdf,dvips]{graphicx}
\usepackage{color}
\usepackage{hyperref}
\usepackage{url}
\usepackage{slashed}
\usepackage{subfigure}
\usepackage[usenames,dvipsnames]{xcolor}
\usepackage{amsmath}
\usepackage{amsfonts}
\usepackage{float} 
\usepackage{amssymb}
\usepackage{epsfig}
\usepackage{graphics}
\usepackage{euscript}
\usepackage{slashed}
\usepackage{epstopdf}
\usepackage[utf8]{inputenc}
\allowdisplaybreaks
\usepackage[normalem]{ulem}
\usepackage{pifont}
\usepackage{dsfont}
\usepackage{MnSymbol}
\usepackage{verbatim}
\usepackage{graphicx}
\usepackage{latexsym}
\usepackage{tikz-feynman}
\usepackage{bbold}

\begin{document}

	\title{{\bf \Large  Investigation of the Bell-CHSH inequality in  diamond regions }}
	
	\vspace{1cm}
	
	\author{M. S.  Guimaraes}\email{msguimaraes@uerj.br} \affiliation{UERJ $–$ Universidade do Estado do Rio de Janeiro,	Instituto de Física $–$ Departamento de Física Teórica $–$ Rua São Francisco Xavier 524, 20550-013, Maracanã, Rio de Janeiro, Brazil}
	
	\author{I. Roditi} \email{roditi@cbpf.br} \affiliation{CBPF $-$ Centro Brasileiro de Pesquisas Físicas, Rua Dr. Xavier Sigaud 150, 22290-180, Rio de Janeiro, Brazil } \affiliation{Institute for Theoretical Physics, ETH Zürich, 8093 Zürich, Switzerland} 
	
	\author{S. P. Sorella} \email{silvio.sorella@fis.uerj.br} \affiliation{UERJ $–$ Universidade do Estado do Rio de Janeiro,	Instituto de Física $–$ Departamento de Física Teórica $–$ Rua São Francisco Xavier 524, 20550-013, Maracanã, Rio de Janeiro, Brazil}

	\begin{abstract}

A numerical setup for the Bell-CHSH inequality for causal diamonds in $1+1$ Minkowski spacetime is presented. Upon choosing a suitable set of test functions supported in the diamonds, sensible violations are reported for the correlation function of Weyl operators of a real scalar massive field  in the vacuum state. 
		
		\end{abstract}

	\maketitle

	\section{Introduction}\label{Sect1}

	The study of the Bell-CHSH inequality \cite{Bell:1964kc,Clauser:1969ny} in relativistic Quantum Field Theory is receiving increasing interest. For instance, the observation \cite{ATLAS:2023fsd} of the entanglement between a top an anti-top quark pair at the ATLAS detector at LHC opens many perspectives to scrutinize the Bell-CHSH inequality at very high energies, see \cite{Barr:2024djo} for a general summary.\\\\Parallel to the experimental and phenomenological efforts, the investigation of the fundamental aspects of the Bell-CHSH inequality in Quantum Field Theory looks rather challenging, requiring the mastering of powerful tools, some examples being: the von Neumann algebras, the modular theory of Tomita-Takesaki and the Haag-Kastler algebraic formulation of Quantum Field Theory, see \cite{Guimaraes:2024mmp} for an updated review. The pivotal theorems obtained by Summers-Werner \cite{Summers:1987fn,Summ,Summers:1987ze} have shown that the Bell-CHSH inequality can be maximally violated already at the level of free fields, underlining the relevance of the entanglement for quantum fields \cite{Witten:2018zxz}. Though, it is safe to state that the concrete implementation of the aforementioned results has not yet been fully accomplished. \\\\In the present work, we pursue the numerical study of the Bell-CHSH inequality started in \cite{Dudal:2023mij,DeFabritiis:2024jfy}, aiming at providing an explicit setup for the detection of possible violations. \\\\More precisely, we shall present a numerical framework for the Bell-CHSH inequality for a real scalar massive field in the vacuum state, by considering causal diamond regions in Minkowski spacetime. The reasons for such an investigation stem from the relevance of the diamonds in Quantum Field Theory \cite{Hislop:1981uh,Brunetti:2002nt,Guido:2008jk}. As we shall see, upon constructing a suitable set of smooth test functions supported in the diamonds, explicit violations of the Bell-CHSH inequality will be reported. \\\\The paper is organized as follows. In Sect.\eqref{SecQuantumWeyl} we outline the main tools which will be employed in the study of the Bell-CHSH correlation function. In Sect.\eqref{tst} we provide the construction of the test functions and we display the numerical results. Sect.\eqref{Cc} is devoted to the conclusion.

	\section{The Bell-CHSH inequality for Weyl operators in the vacuum states}\label{SecQuantumWeyl}
	
\begin{figure}[t!]
	\begin{minipage}[b]{0.6\linewidth}
		\includegraphics[width=\textwidth]{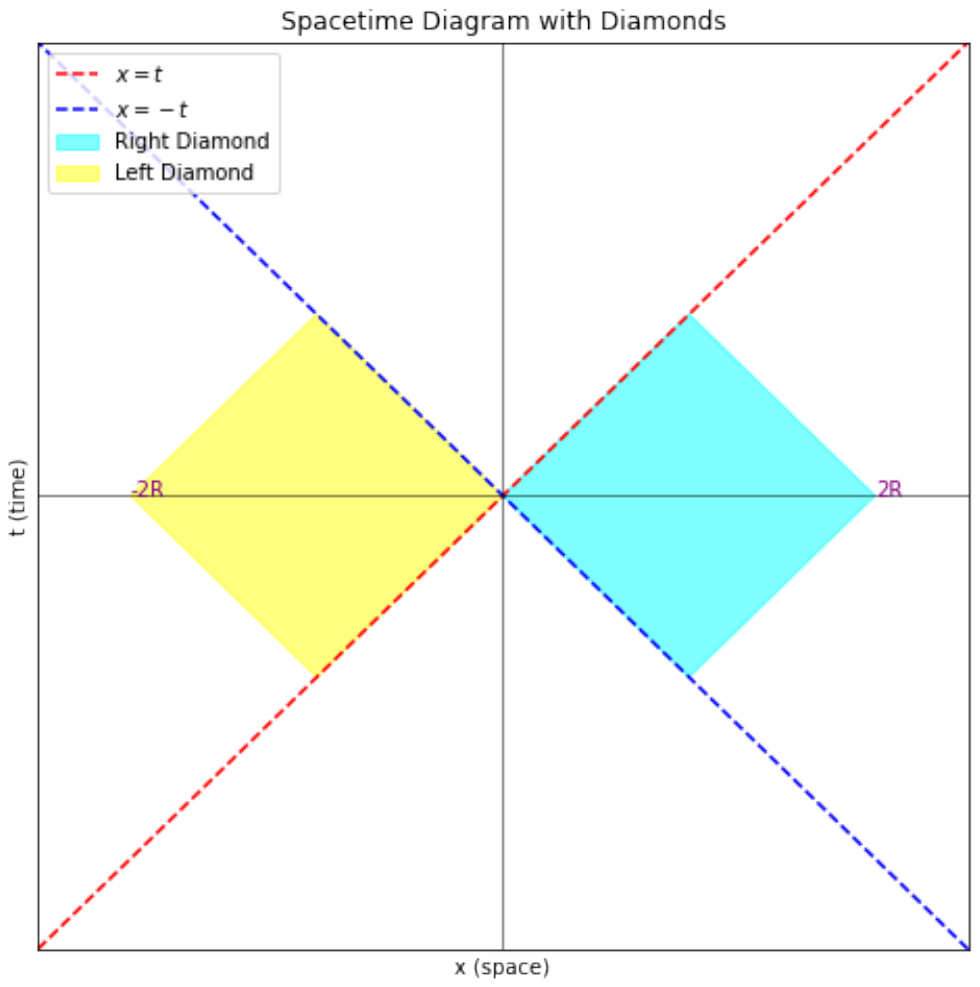}
	\end{minipage} \hfill
\caption{Causal diamond regions. The test functions $(f,f')$ are supported in  right diamonds while $(g,g')$ in  left ones.  }
	\label{diamonds}
	\end{figure}

	We start by considering a free massive real scalar field in $1+1$ Minkowski spacetime, namely 
	\begin{equation} \label{qf}
		\phi(t,x) = \int \! \frac{d k}{2 \pi} \frac{1}{2 \omega_k} \left( e^{-ik_\mu x^\mu} a_k + e^{ik_\mu x^\mu} a^{\dagger}_k \right), 
	\end{equation} 
	where $\omega_k  = k^0 = \sqrt{k^2 + m^2}$. The canonical commutation relations read
	\begin{align}
		[a_k, a^{\dagger}_q] &= 2\pi \, 2\omega_k \, \delta(k - q), \\ \nonumber 
		[a_k, a_q] &= [a^{\dagger}_k, a^{\dagger}_q] = 0. 
	\end{align}
	It is well established that quantum fields are operator-valued distributions \cite{Haag:1992hx}. As such, they  have to be smeared out to yield  well-defined operators acting on the Hilbert space, {\it i.e.}
	\begin{align} 
		\phi(h) = \int \! d^2x \; \phi(x) h(x),
	\end{align}
	where $h$ is a real smooth test function with compact support, $h \in \mathcal {C}_{0}^{\infty}(\mathbb{R}^4)$. Using the smeared fields, the Lorentz-invariant inner product is introduced by means of the two-point  smeared Wightman function: 
	\begin{align} \label{InnerProduct}
		\langle f \vert g \rangle &= \langle 0 \vert \phi(f) \phi(g) \vert 0 \rangle =  \frac{i}{2} \Delta_{PJ}(f,g) +  H(f,g)
	\end{align}
	where $ \Delta_{PJ}(f,g)$ and $H(f,g)$ are the smeared versions of the Pauli-Jordan and Hadamard expressions:
	\begin{align}
		\Delta_{PJ}(f,g) &=  \int \! d^2x d^2y f(x) \Delta_{PJ}(x-y) g(y) \;, \\
		H(f,g) &=  \int \! d^2x d^2y f(x) H(x-y) g(y)\;, \label{mint}
	\end{align}
with $\Delta_{PJ}(x-y)$ and $H(x-y)$ given by
\begin{eqnarray} 
\Delta_{PJ}(t,x) & =&  -\frac{1}{2}\;{\rm sign}(t) \; \theta \left( \lambda(t,x) \right) \;J_0 \left(m\sqrt{\lambda(t,x)}\right) \;, \nonumber \\
H(t,x) & = & -\frac{1}{2}\; \theta \left(\lambda(t,x) \right )\; Y_0 \left(m\sqrt{\lambda(t,x)}\right)+ \frac{1}{\pi}\;  \theta \left(-\lambda(t,x) \right)\; K_0\left(m\sqrt{-\lambda(t,x)}\right) \;, \label{PJH}
\end{eqnarray}
where 
\begin{equation} 
\lambda(t,x) = t^2-x^2 \;, \label{ltx}
\end{equation}
and $(J_0,Y_0,K_0)$ are Bessel functions, while $m$ is the mass parameter. \\\\Both Hadamard and Pauli-Jordan distributions are Lorentz-invariant. Moreover, $\Delta_{PJ}(x)$  encodes the information on relativistic causality, being vanishing  outside of the light cone.   Also,  $\Delta_{PJ} (x)$ and $H(x)$ are respectively odd and even under the change $x \rightarrow -x$. Using the smeared fields, we can write $\left[\phi(f), \phi(g)\right] = i \Delta_{PJ}(f,g)$, allowing us to recast causality in this setting as having $\left[\phi(f), \phi(g)\right] = 0$ when the supports of $f$ and $g$ are space-like separated. \\\\In order to introduce the Bell-CHSH correlation function, we shall employ the unitary Weyl operators  \cite{DeFabritiis:2023tkh} 
\begin{align}
		A_f = e^{i \phi(f)} \;, \qquad A_f^{\dagger} A_f = A_f A_f^\dagger = 1 \;., \qquad A^{\dagger}_f = A_{-f} \;.  \label{ww}
	\end{align}
These operators obey the Weyl algebra: 
	\begin{align}\label{WeylAlgebra}
		A_f \,  A_g = e^{-\frac{i}{2} \Delta(f,g)} \,A_{(f+g)} \;. 
	\end{align}
When the supports of $f$ and $g$ are space-like separated, the Pauli-Jordan function vanishes, ensuring that $A_f A_g = A_{(f+g)}$.  Computing the vacuum expectation value of the Weyl operator, one finds:
	\begin{align}\label{WeylVEV}
		\langle 0 \vert A_f \vert 0 \rangle = \langle 0 \vert A_{-f} \vert 0 \rangle = e^{-\frac{1}{2} \vert\vert f \vert\vert^2},
	\end{align}
	where  $\vert\vert f \vert\vert^2 = \langle f \vert f \rangle$. In particular, when $(f,g)$ are space-like, 
	\begin{equation} 
	\langle 0 |\; A_f A_g \;|0 \rangle = \langle 0 |\; A_{f+g}  \;|0 \rangle = e^{-\frac{1}{2} \vert\vert f +g \vert\vert^2} \;. \label{usw}
	\end{equation}
To investigate the Bell-CHSH inequality, we follow \cite{Guimaraes:2024mmp,DeFabritiis:2023tkh,Guimaraes:2024alk,Guimaraes:2024lqf} and consider the Bell-CHSH type expression 
	\begin{align}
		\langle 0 \vert \mathcal{C} \vert 0 \rangle &= 	\langle 0 \vert \left(A_f + A_{f'}\right) A_g + \left(A_f - A_{f'}\right) A_{g'} \vert 0 \rangle  \;, \label{Bll}
	\end{align}
where $(f,f')$  and $(g,g')$ are pairs of space-like test functions. More precisely, $(f,f')$ will be supported in  diamond regions located in the right wedge ${\cal W}_R$, while $(g,g')$ will be supported in  diamonds located in the left wedge ${\cal W}_L$\footnote{The regions ${\cal W}_R$ and ${\cal W}_L$ are defined by 
\begin{equation} 
{\cal W}_R = \big\{ (x,t) \in {\mathbb{R}^2}, \; x > |t| \; \big\}\;, \qquad {\cal W}_L = \big\{ (x,t) \in {\mathbb{R}^2}, \; -x > |t| \; \big\}\;.  \label{rdw2}
\end{equation} }, as illustrated in Fig.\eqref{diamonds} 

The correlation function \eqref{Bll} violates the Bell-CHSH inequality whenever 
\begin{equation} 
	2 < | \langle 0 \vert \mathcal{C} \vert 0 \rangle | \le 2 \sqrt{2} \;. \label{viol}
\end{equation}
Since $(f,f')$ and $(g,g')$ are space-like, equation  \eqref{Bll}
 can be expressed entirely in terms of the Hadamard distribution. In fact, from eq.\eqref{usw} and from 
 \begin{align}\label{NormaFG}
			\vert\vert f+g \vert\vert^2 = H(f,f) + 2 H(f,g) + H(g,g) \;, 
	\end{align}
it follows that 
\begin{align}\label{BellFG}
		\langle 0 \vert \mathcal{C} \vert 0 \rangle &= e^{-\frac{1}{2}\left[H(f,f) + 2 H(f,g) + H(g,g)\right]} 
		+ e^{-\frac{1}{2} \left[H(f',f') + 2 H(f',g) + H(g,g)\right]} \nonumber \\
		&+ e^{-\frac{1}{2} \left[H(f,f) + 2 H(f,g') + H(g',g')\right]}  
		- e^{-\frac{1}{2} \left[H(f',f') + 2 H(f',g') + H(g',g')\right]} \;.	
\end{align}

\section{Choice of the test functions, numerical integration and results}\label{tst}

We proceed by specifying the shape of the test functions which will be employed. Let us consider first the right diamond, specified by the condition 
\begin{equation} 
|x-r| + |t| \le r \;. \label{rd}
\end{equation} 
 For $(f,f')$ we write 
\begin{align}
f(t,x) = \eta
\left\{
    \begin {aligned}
         & e^{-\frac{a}{r^2 - (|x-r| +|t|)^2}} \quad & |x-r| + |t| \le r   \\
         & 0 \quad & {\rm elsewhere}                   
    \end{aligned}
\right. \label{ff}
\end{align}
and 
\begin{align}
f'(t,x) = \eta'
      \left\{
    \begin {aligned}
         & e^{-\frac{a}{r'^2 - (|x-r'| +|t|)^2}} \quad & |x-r'| + |t| \le r'   \\
         & 0 \quad & {\rm elsewhere}                   
    \end{aligned}
\right. \label{ffp}
\end{align}
where $(a,a',r,r',\eta,\eta)$ are arbitrary parameters, to be fixed at the best convenience. 
Analogously, in the left diamonds, one  considers 
\begin{align}
g(t,x) = \sigma
\left\{
    \begin {aligned}
         & e^{-\frac{b}{r^2 - (|x+r| +|t|)^2}} \quad & |x+r| + |t| \le r   \\
         & 0 \quad & {\rm elsewhere}                   
    \end{aligned}
\right. \label{gg}
\end{align}
and 
\begin{align}
g'(t,x) = \sigma'
      \left\{
    \begin {aligned}
         & e^{-\frac{b'}{r'^2 - (|x+r'| +|t|)^2}} \quad & |x+r'| + |t| \le r'   \\
         & 0 \quad & {\rm elsewhere}                   
    \end{aligned}
\right. \label{ggp}
\end{align}

with $(b,b',\sigma,\sigma')$ free parameters. The behavior of $f$ and $g$ is shown in Figs.\eqref{ff},\eqref{gg}. One sees that $f$ vanishes in the left wedge, while $g$ vanishes in the right one. 
\begin{figure}[t!]
	\begin{minipage}[b]{0.4\linewidth}
		\includegraphics[width=\textwidth]{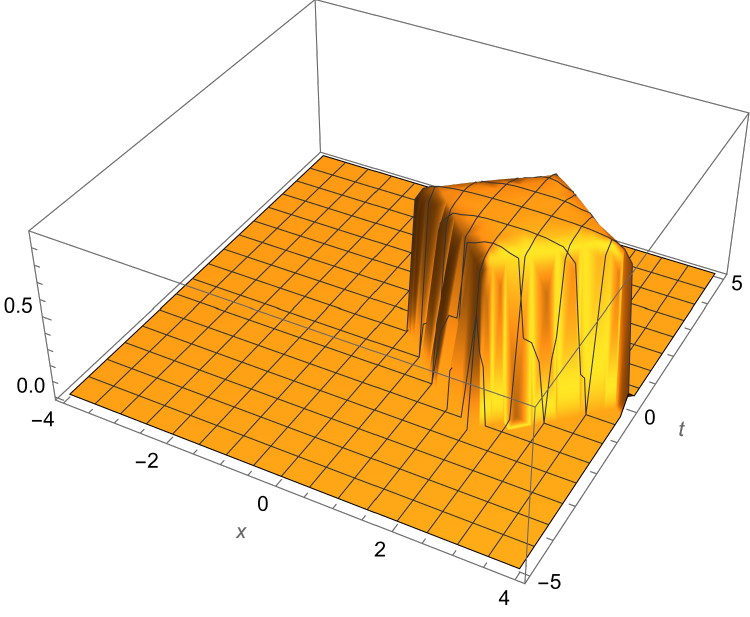}
	\end{minipage} \hfill
\caption{Plot of the test function $f(t,x)$, for $(a=0.1, \eta=1, r=2 )$.   }
	\label{ff}
	\end{figure}
	
\begin{figure}[t!]
	\begin{minipage}[b]{0.4\linewidth}
		\includegraphics[width=\textwidth]{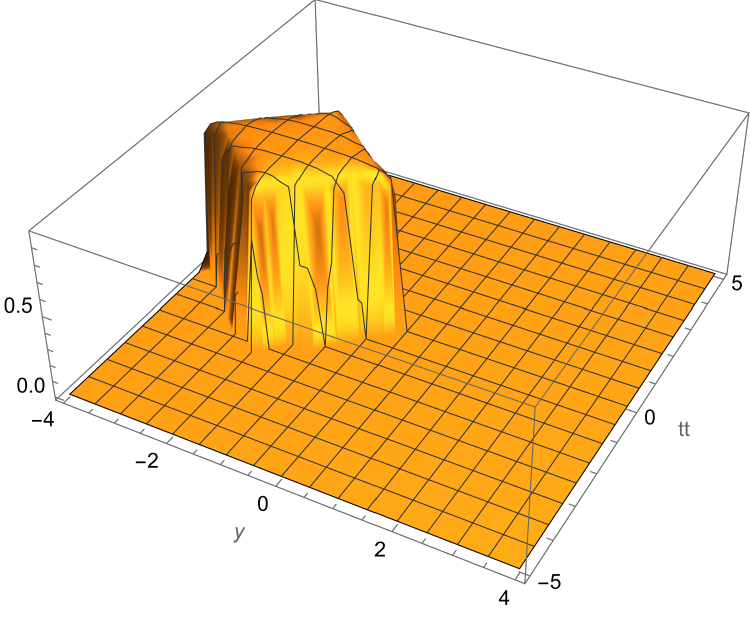}
	\end{minipage} \hfill
\caption{Plot of the test function $g(t,x)$, for $(b=0.1, \sigma=1, r=2 )$.   }
	\label{gg}
	\end{figure}

\vspace{0.5cm}
\noindent Concerning the details of the numerical integration of the scalar products appearing in eq.\eqref{BellFG}, the typical integral is shown in equation \eqref{mint}. Due to the difficulties of evaluating the Fourier transformation of the test functions in momentum space in closed analytic form, expression \eqref{mint} has been evaluated as it  stands, {\it i.e.} in configuration space. We have employed Mathematica, relying on two methods of integration: QuasiMonteCarlo and MultidimensionalRule. The parameters $(a,\eta, a', \alpha',\eta')$, $(b,\sigma, b',\sigma')$ as well as the mass $m$ and $(r,r') $ have been selected by performing tests using  an aleatory algorithm. In each test,  $10^5$ aleatory values for the parameters have been checked. \\\\The following table gives an idea of the results which have been obtained. One sees that the test functions \eqref{ff}-\eqref{ggp} give rise to  nice violations of the Bell-CHSH inequality.  
 



 \begin{table}[h!]
		\begin{tabular}[t]{|c|c|c|c|c|c|c|c|c|c|c|c| }
			\hline
	$a$ & $\eta$ & $b$ & $\sigma$ & $a'$ & $\eta'$ & $b'$ & $\sigma'$ &	$m$  & $r$ & $r'$ & 
			$\langle \mathcal{C} \rangle$  \\
			\hline 
			
	 0.453107 &  0.06256&
 0.241230&    0.033623 &
 3.008120 &   4.486029 &
  0.699209 &   4.096952 &
  0.009390 & r = 1.859616 &
 0.840575&
 2.067 	\\	
			\hline
			
 0.394546& 0.170251 & 
 0.446326 & 0.080372& 
  4.930245 & , 4.697725&  
  3.659304 &12.398896 &  
  0.000193&1.196910& 
  1.192083& 
2.071 \\
\hline 
	0.693921 &0.28891 &   0.299133 &  0.082498 &  
  4.6219, &  0.397221 &1.44436 &  4.03835 & 
 0.000359 & 1.34912 & 1.06941 & 
\ 2.103`			 \\
			\hline
		\end{tabular}
		\caption{Results obtained for the Bell-CHSH correlation function \eqref{Bll}. The values of the violation are reported in the last column.}
		\label{tabelaMax}
	\end{table}	
 
It is instructive to compare the results obtained with diamonds with those reported in \cite{Guimaraes:2024mmp} for wedge regions, see Table \eqref{tw} . In both cases the same numerical setup has been employed. 
One sees that the use of diamond regions yields slightly better results. Besides, the running time for the employed algorithms turns out to be faster than that for wedges regions.

 \begin{table}[h!]
		\begin{tabular}[t]{|c|c|}
			\hline
		$m$  &
			$\langle \mathcal{C} \rangle$  \\
			\hline 
0,00251 & 2.034 \\
\hline 			
	
 0.00027	& 2.044\\	
			\hline

		\end{tabular}
		\caption{Results obtained for the Bell-CHSH correlation function \eqref{Bll} in wedge regions  \cite{Guimaraes:2024mmp}.}
		\label{tw}
	\end{table}

\section{Conclusions}\label{Cc}

In this work we have pursued the numerical investigation of the violations of the Bell-CHSH inequality which may occur in the correlation function of Weyl operators in the case of a real massive scalar field in $1+1$ Minkowski spacetime. \\\\Causal diamond regions have been considered. Upon constructing a set of suitable test functions enabling to localize the Weyl operators in the corresponding diamonds, sensible violations of the Bell-CHSH inequality have been found, as summarized in Table \eqref{tabelaMax}. \\\\Let us also add that, comparing the present results with those obtained previously for the scalar field \cite{Guimaraes:2024mmp}, the use of diamond regions seems very suited for a numerical investigation, due to the convergence properties of the configuration integrals, eq.\eqref{mint}, resulting in good numerics and fast algorithms.

\section*{Acknowledgments}
 I.R.  acknowledge the warm hospitality of the QIT group in ETHz where this work has been completed. The authors would like to thank the Brazilian agencies CNPq, CAPES end FAPERJ for financial support.  S.P.~Sorella, I.~Roditi, and M.S.~Guimaraes are CNPq researchers under contracts 301030/2019-7, 311876/2021-8, and 309793/2023-8, respectively. 




\end{document}